\documentstyle[12pt]{article}
\begin{document} \pagestyle{empty} \large \noindent
G\"{o}teborg ITP 92-49\\ November 1992\\
\begin{center} \LARGE \bf \vspace*{40mm}

Unification of gravity and Yang-Mills theory in (2+1)-dimensions.\\
\vspace*{20mm} \Large \bf
Peter Peld\'{a}n\\
\vspace*{5mm} \large
Institute of Theoretical Physics\\
S-412 96  G\"{o}teborg, Sweden\\
Internet: tfepp@fy.chalmers.se\\
\vspace*{30mm} \Large \bf
Abstract\\ \end{center} \normalsize
A gauge and diffeomorphism invariant theory in (2+1)-dimensions is presented in
both
first and second order Lagrangian form as well as in a Hamiltonian form. For
gauge
group $SO(1,2)$, the theory is shown to describe ordinary Einstein gravity with
a
cosmological constant. With gauge group $G^{tot}=SO(1,2)\otimes G^{YM}$, it is
shown
that the equations of motion for the $G^{YM}$ fields are the Yang-Mills
equations. It
is also shown that for weak $G^{YM}$ Yang-Mills fields, this theory agrees with
the
conventional Einstein-Yang-Mills theory to lowest order in Yang-Mills fields.

Explicit static and rotation symmetric solutions to the Einstein-Maxwell theory
are
studied both for the conventional coupling and for this unified theory. In the
electric solution to the unified theory, point charges are not allowed, the
charges
must have spatial extensions.\\


\newpage \pagestyle{plain} \section{Introduction}

The quest for a unified theory of gravity and Yang-Mills theory is even older
than the
theory of general relativity and Yang-Mills theory themselves. It started
already
in 1914
with Nordstr\"{o}m's work of unifying his scalar theory of gravitation with
Maxwell's
theory of electro-magnetism \cite{Nordstrom}. This attempt was in 1921 followed
by
Kaluza's five dimensional Einstein equation, which was shown to describe the
coupled
Einstein-Maxwell theory \cite{Kaluza}. This formulation is normally called the
Kaluza-Klein theory, since Klein rediscovered Kaluza's theory in 1926
\cite{Klein}.
Much later, after the invention of Yang-Mills theory in 1954, a Kaluza-Klein
unification
of gravity and Yang-Mills theories was considered by DeWitt, Trautman, Kerner
and
others \cite{Mod.Kaluza.Klein}.

The common idea behind all these attempts is that space-time has some extra
space-like
dimensions besides the normal (3+1) observable ones. These extra dimensions
should
then for some reason be compactified on a very small length-scale, and
therefore be
non-observable at "normal" energy-scales.

What I will present here in this paper, is a somewhat different idea of how to
find a
unified theory of gravity and Yang-Mills theory. Instead of enlarging the
space-time,
I consider an enlarged internal symmetry group. The normal "internal" symmetry
group
for gravity in (2+1)-dimensions is the Lorentz group $SO(1,2)$. I study a
theory
valid for an arbitrary gauge group, which reduces to the conventional Einstein
theory
for gravity if one chooses the gauge group to be $SO(1,2)$. For other gauge
groups,
like $G^{tot}=SO(1,2)\otimes G^{YM}$, the theory has an interpretation of
gravity
coupled to Yang-Mills theory. (Weinberg \cite{Weinberg} has considered a
related
generalization of gravity, in which he enlarges both the space-time dimensions
and the
internal symmetry group.)

In section 2, the first order Lagrangian for the unified theory is given, and
it is
shown that for $SO(1,2)$ the theory reduces to Einstein gravity with a
cosmological
constant. The problem with the first order formulation is that there exist no
obvious
metric-definition, meaning that the physical interpretation is unclear.

In section 3, the Hamiltonian formulation is given, and a constraint analysis
is
performed. For a canonical formulation there exist a prescription, based on
purely
geometrical considerations,
 of how to
identify the metric from the constraint algebra in any diffeomorphism invariant
theory.
I use this prescription, and the metric is identified. Then, I
compare the unified theory to the conventional Einstein-Yang-Mills theory, and
show
that in the weak field limit, the two formulations agree. It is also shown that
the
equations of motion governing the Yang-Mills fields is the normal Yang-Mills
equations
even for an arbitrary strength of the Yang-Mills field.

In section 4, the second order pure connection formulation of the unified
theory is
presented, as well as the metric formulas for the metric in terms of the
connection.

Finally in section 5, explicit static and rotation symmetric solutions to both
the
conventional and the unified Einstein-Maxwell theory, is found and compared. It
is
shown that for weak Maxwell fields the solutions to the conventional theory can
be
found as the lowest order
terms in the solutions to the unified theory. One new interesting feature is
found in
the electric solution to the unified theory: Point charges are not allowed,
there exist
no solutions inside a radius $r=q$ in Schwardschild coordinates.

\section{First order Lagrangian formulation}
In this section, I will present a first order Lagrangian, which is a function
of a gauge
connection and a trio of Lie-algebra valued vector fields.
For the special choice of $SO(1,2)$ as
the gauge group, this Lagrangian describes ordinary Einstein gravity with a
cosmological
constant. In that case, the gauge connection is the spin-connection, and the
Lie-algebra
valued vector fields equals the triad field. For other gauge groups, there is
no obvious
interpretation of the theory, here in the first order Lagrangian form. However,
in the
Hamiltonian formulation of the theory, which is given in section 3, there
exists a
geometrical prescription of how to read off the metric in the constraint
algebra, and it
will then become clear that the metric in this theory is still the square of
the vector
fields.

The Lagrangian, valid for an arbitrary gauge group is

\begin{equation} {\cal L}= \epsilon^{\alpha \beta \gamma} e_{\alpha I}
 F^I_{\beta \gamma}
+ \lambda \sqrt{-g} \label{2.1} \end{equation}

where $e_{\alpha I}$ and $A_\alpha ^I$ are the basic fields. $F^I_{\alpha
\beta}=
\partial_{\alpha} A_{\beta}^I - \partial_{ \beta} A_{\alpha}^I +
f^{IJK} A_{\alpha J} A_{\beta K}$. $f^{IJK}$ are the structure constants of the
gauge
group, and the "gauge-indices" are raised and lowered with a bilinear invariant
form
of the Lie-algebra. For $SO(1,2)$ I will always choose this "group-metric" to
be
$\eta_{IJ}=diag(-1,1,1)$.
The space-time indices are denoted $\alpha, \beta, \gamma$ and take
values $0,1,2$. $g_{\alpha \beta}:=e_{\alpha I} e_{\beta}^I$ and $g=\frac{1}{6}
\epsilon^{\alpha \beta \gamma} \epsilon^{\delta \epsilon \sigma} g_{\alpha
\delta}
g_{\beta \epsilon} g_{\gamma \sigma}$. $\epsilon^{\alpha \beta \gamma}$ is
totally
anti-symmetric and $\epsilon^{012}=1$ in every coordinate system, which means
that
$\epsilon^{\alpha \beta \gamma}$ is a tensor density of weight $+1$.

Now, for gauge group $SO(1,2)$ it is true that $g=-e^2$, where $e=\frac{1}{6}
\epsilon^{\alpha \beta \gamma} f^{IJK} e_{\alpha I} e_{\beta J} e_{\gamma K}$,
and the
Lagrangian (\ref{2.1}) can be rewritten as

\begin{equation} {\cal L}_1= \epsilon^{\alpha \beta \gamma} e_{\alpha I}
 F^I_{\beta \gamma}
+ \lambda e \label{2.2} \end{equation}

which is a Lagrangian for Einstein gravity with a cosmological constant. See
Witten
\cite{Witt.2+1}.
Using another formula valid for $SO(1,2)$: $e_{\alpha I} \epsilon^{\alpha \beta
\gamma}=e f_{IJK} e^{\beta J} e^{\gamma K}$, where $e^{\beta J}$ is the inverse
to
$ e_{\alpha I}$, the Lagrangian can be changed into:

\begin{equation} {\cal L}_2= e f_{IJK} e^{\beta J} e^{\gamma K}F^I_{\beta
\gamma}
+ \lambda e \label{2.3} \end{equation}

which is the ordinary Hilbert-Palatini Lagrangian. This shows that the
Lagrangian
(\ref{2.1}) is a natural generalization of the Hilbert-Palatini Lagrangian,
to other gauge groups. It is also possible to generalize (\ref{2.2}) and
(\ref{2.3})
directly to other gauge groups. This will however not lead to any interesting
Hamiltonians. ${\cal L}_2$ will lead to second class constraints, if the gauge
group
has dimension greater than three, and ${\cal L}_1$ will lead to a trivial
Hamiltonian
describing a theory without any local degrees of freedom. So, it is only
(\ref{2.1})
that will give an interesting Hamiltonian.

Returning to (\ref{2.1}), the equations of motion is

\begin{eqnarray}
\frac{\delta S}{\delta A_{\alpha I}}&=&-2 {\cal D}_{\beta} \epsilon^{\beta
\alpha
\gamma} e_\gamma ^I=0 \label{2.4} \\
\frac{\delta S}{\delta e_{\alpha I}}&=&\epsilon^{\alpha \beta \gamma}
F^I_{\beta
\gamma} + \lambda \sqrt{-g}e^{\alpha I}=0 \label{2.5} \end{eqnarray}

where $e^{\alpha I}:=g^{\alpha \beta} e^I_\beta$ and $g^{\alpha \beta}$ is the
inverse
to $g_{\alpha \beta}$. Note that it is true that $e^{\alpha I} e_{\beta
I}=\delta^\alpha _\beta$ for an arbitrary gauge group, but it is not true in
general
that
$e^{\alpha I} e_{\alpha J}=\delta^I _J$. It is only for three dimensional gauge
groups
that both relations hold. For higher dimensional groups,  $e^{\alpha I}
e_{\alpha J}$ is
just a degenerate matrix of rank three.

Equation (\ref{2.4}) is normally, for $SO(1,2)$, called the torsion-free
condition,
which can be solved to give the spin-connection as a function of the triad. For
higher
dimensional gauge groups, it is still possible to solve (\ref{2.4}) to get an
expression for $A_{\alpha I}$ in terms of $e_{\alpha I}$. This will however not
totally fix the connection. The solution of (\ref{2.4}) will only give the
parts of
the connection that are non-orthogonal to $e_{\alpha I}$ in the
"gauge-indices".
Equation (\ref{2.5}) is for $SO(1,2)$ just Einstein's equations with a
cosmological
constant, but for other gauge groups, I see no obvious interpretation. It's
interpretation will become much more transparent in the Hamiltonian
formulation, in
next section.

\section{Hamiltonian formulation}
Starting from the Lagrangian (\ref{2.1}), valid for an arbitrary gauge group, I
will
now perform the Legendre transform to a Hamiltonian formulation, find all
constraints
and calculate the constraint algebra. Then, this Hamiltonian formulation will
be
compared to the conventional Ashtekar and Witten Hamiltonian formulations of
(2+1)-dimensional Einstein gravity. See Bengtsson \cite{Beng.2+1}, and Witten
\cite{Witt.2+1}. Using the constraint algebra, I will then read off the
space-time
metric in this theory, and finally I will compare this generalized theory to
the
normal Einstein-Yang-Mills theory, to see whether it is possible to find the
normal
theory in some limit of the unified theory.

\subsection{Legendre transform}
Inspection of the Lagrangian (\ref{2.1}) gives that the velocities of $e_{0I}$
and
$A_{0I}$ are absent in the Lagrangian, meaning that these fields will become
Lagrange
multiplier fields. So the only field that will have a non-vanishing momenta is
$A_{a
I}$. Defining the momenta

\begin{equation} \Pi^{aI}:=\frac{\partial {\cal L}}{\partial
\dot{A}_{aI}}=\epsilon^{ab}e^I_b \label{3.2} \end{equation}

where $a,b,c$ are spatial indices, taking values $1$ and $2$. $\epsilon^{ab}=
\epsilon^{0ab}$. I also define $\epsilon_{ab}$ to be anti-symmetric and
$\epsilon_{12}=1$. This means that $\epsilon^{ab} \epsilon_{cd}=\delta^a_c
\delta^b_d
- \delta^b_c \delta^a _d$. Now, (\ref{3.2}) can be easily inverted, to get

\begin{equation} e_{aI}=\epsilon_{ba} \Pi^b_I \label{3.3} \end{equation}

Putting this into the Lagrangian (\ref{2.1}) gives

\begin{equation} {\cal L}= \Pi^a_I F^I_{0a} + e_{0I} \Psi^I + \lambda
\sqrt{-e_{0I}
e_0^I det(\Pi^{aI} \Pi^b _I) + \epsilon_{ab} \epsilon_{cd} \Pi^{bI} \Pi^d_I
\Pi^{aJ}
e_{0J} \Pi^{cK} e_{0K}} \label{3.5} \end{equation}

where $\Psi^I:=\epsilon^{ab} F^I_{ab}$ and $det( \Pi^{aI} \Pi^b _I)=\frac{1}{2}
\epsilon_{ab} \epsilon_{cd} \Pi^{bI} \Pi^d_I \Pi^{aJ} \Pi^c_J$.

The Lagrangian has an inhomogeneous dependence of the Lagrange multiplier field
$e_{0I}$, which will lead to constraints that depend on the Lagrange multiplier
field.
This is normally not wanted, so I eliminate this inhomogeneous dependence
first. To do
this, I project out two components of $e_{0I}$:

\begin{equation} e_{0I}=M_a \Pi^a_I + V_I,\hspace{5mm} V_I \Pi^{aI}=0
\Rightarrow M_a=e_{0I} \Pi^{bI} q_{ba}, \hspace{5mm}V_I=e_{0I}-M_a \Pi^a_I
\label{3.7}
 \end{equation}

where $q_{ab}$ is the inverse to $q^{ab}:=\Pi^{aI} \Pi^b_I$.

This gives the Lagrangian

\begin{equation} {\cal L}=\Pi^a_I F^I_{0a} + M_a \Pi^a_I \Psi^I + V_I \Psi^I +
\lambda \sqrt{- V^I V_I det(q^{ab})} \label{3.8} \end{equation}

The Lagrangian is still inhomogeneous in $V^I$, but the variation of $V^I$
gives:

\begin{equation} \frac{\delta S}{\delta V^I}= \Psi_I - \lambda
\frac{1}{\sqrt{-V^I V_I
det(q^{ab})}} det(q^{ab}) V_I=0 \label{3.9} \end{equation}

and taking the square of (\ref{3.9}),

\begin{equation} \Psi^I \Psi_I=-\lambda^2 det(q^{ab}) \label{3.95}
\end{equation}

This means that the variation of $V^I$ imposes the constraint (\ref{3.95}), and
I can
remove all terms in the Lagrangian containing $V^I$, and put in the equivalent
constraint (\ref{3.95}) with a Lagrange multiplier, $N$. Then finally, I
redefine the
Lagrange multiplier $N^a:=2 \epsilon^{ba} M_b$ for later comparison with the
normal
formulations of gravity. The Hamiltonian is ${\cal H}=\Pi^{aI}
\dot{A}_{aI}-{\cal L}$

\begin{eqnarray}
{\cal H}_{tot}&=&N {\cal H} + N^a {\cal H}_a - A_{0I} {\cal G}^I \label{3.11}\\
{\cal H}&:=&\frac{1}{4 \lambda} (\Psi^I \Psi_I + \lambda^2 det(q^{ab})) \approx
0
\nonumber\\
{\cal H}_a&:=&\frac{1}{2}\epsilon_{ab} \Pi^b_I \Psi^I \approx 0 \nonumber\\
{\cal G}^I&:=&{\cal D}_a \Pi^{aI}=\partial_a \Pi^{aI} + f^{IJK} A_{aJ} \Pi^a_K
\approx 0
\nonumber \end{eqnarray}

where $N, N^a, A_{0I}$ are Lagrange multiplier fields, and ${\cal H}, {\cal
H}_a$ and
${\cal G}_I$ are constraints following from variation of the Lagrange
multiplier
fields. The constraints are normally called the Hamiltonian constraint, the
vector
constraint and Gauss' law.

This Hamiltonian will for now on be referred to as "the unified Hamiltonian"
and the
theory it describes for an arbitrary gauge group will be called "the unified
theory".
In subsection 3.4, it will be shown how this unified Hamiltonian is related to
the
Ashtekar and Witten Hamiltonians for pure gravity.

 \subsection{Constraint analysis}
Now, to make sure that there are no more constraints in the
theory, one must check that the time evolution of the constraints vanishes
weakly. And
to check that, one needs the constraint algebra. To derive the constraint
algebra, I
start by deriving how the basic canonical fields transforms under
transformations
generated by ${\cal G}_I$ and $\tilde{{\cal H}}_a:={\cal H}_a + A_{aI}{\cal
G}^I$:
\begin{eqnarray}
\delta^{{\cal G}^I [\Lambda_I]} \Pi^a_I (x)=\{\Pi^a_I (x),{\cal
G}_I[\Lambda^I]\}&=&\Lambda^J (x) \Pi^{aK}(x) f_{JKI} \label{3.121}\\
\delta^{{\cal G}^I [\Lambda_I]}A_{aI}(x)=\{A_{aI}(x),{\cal G}_I[\Lambda^I]\}
&=&-{\cal D}_a \Lambda_I (x) \label{3.122}\\
\delta^{\tilde{{\cal H}}_a[N^a]}\Pi^a_I(x)=\{\Pi^a_I (x),{\cal
H}_a[N^a]\}&=&-\pounds_{N^a}\Pi^a_I (x) \label{3.131} \\
\delta^{\tilde{\cal H}_a[N^a]} A_{aI}(x)=\{A_{aI}(x),{\cal H}_a[N^a]\}
&=&-\pounds_{N^a} A_{aI}(x) \label{3.132}
\end{eqnarray}

where ${\cal G}^I [\Lambda_I]=\int d^2 x {\cal G}^I(x) \Lambda_I(x)$, and
$\pounds_{N^a}$ is the Lie-derivative in the direction $N^a$. This means that
${\cal G}^I$ is the generator of gauge transformations, and $\tilde{{\cal
H}}_a$ is
the generator of spatial diffeomorphisms. It is then clear how an arbitrary,
gauge
and diffeomorphism covariant, function $\Phi_{aI} (\Pi^a_I,A_b^J)$
of the basic fields will transform under the
transformations generated by ${\cal G}^I$ and $\tilde{{\cal H}}_a$.

\begin{eqnarray}
\delta^{{\cal G}^I [\Lambda_I]} \Phi_{aI} (\Pi^a_I,A_b^J)&=&\Lambda^J
\Phi_a^K(\Pi^a_I,A_b^J) f_{JKI}\label{3.14} \\
\delta^{\tilde{{\cal H}}_a[N^a]} \Phi_{aI} (\Pi^a_I,A_b^J)&=&-\pounds_{N^a}
\Phi_{aI}(\Pi^a_I,A_b^J) \end{eqnarray}

This makes it simple to calculate the constraint algebra

\begin{eqnarray}
\{{\cal G}^I [\Lambda_I],{\cal G}^J[\Gamma_J]\}&=&{\cal G}^K[f_{KIJ}\Lambda^I
\Gamma^J]\label{3.15} \\
\{{\cal G}^I [\Lambda_I],\tilde{{\cal H}}_a[N^a]\}&=&{\cal
G}^I[\pounds_{N^a}\Lambda^I]\\
\{{\cal G}^I [\Lambda_I],{\cal H}[N]\}&=&0\\
\{\tilde{{\cal H}}_a[N^a],\tilde{{\cal H}}_b[M^b]\}&=&\tilde{{\cal
H}}_a[\pounds_{M^b}
N^a]\\
\{{\cal H}[N],\tilde{{\cal H}}_a[N^a]\}&=&{\cal H}[\pounds_{N^a}N]
\end{eqnarray}

The reason why the Poisson-bracket between ${\cal G}^I$ and $\tilde{{\cal
H}}_a$ is
not zero is that $\tilde{{\cal H}}_a$ has a non-gauge covariant dependence on
$A_{aI}$.
Now, it is only one Poisson-bracket left to calculate:

\begin{equation} \{{\cal H}[N],{\cal H}[M]\}={\cal H}_a[\Pi^{aI} \Pi^b_I
(N\partial_b
M - M\partial_b N)] \label{3.16} \end{equation}

Equations (\ref{3.15})-(\ref{3.16}) then give the complete constraint algebra,
and the
constraints form a first class set, and therefore the time-evolution of the
constraints is automatically vanishing on the constraint surface. So the
Hamiltonian
(\ref{3.11}) is complete, and defines a consistent theory, in this sense.

\subsection{Reading off the metric}
Given a diffeomorphism invariant theory with no obvious physical
interpretation, it
is not a trivial task to identify the metric-field in the theory.
 A basic requirement of the metric-definition should be that a given
test-particle
 should propagate on geodesics to that metric.
 And
if the theory is physically sensible, the propagation should be causal. Hojman
Kucha\v{r} and Teitelboim \cite{HKT} has examined a related question. They
considered
a general canonical formulation of a diffeomorphism invariant theory, in which
they
defined the generator of parallel deformations ${\cal H}_\|$
, and the generator of normal deformations ${\cal H}_\perp$
with respect to the hypersurface. Then, by requiring path-independence of
deformations, they derived a specific deformation algebra:
$\{{\cal H}_\|,{\cal H}_\|\}\sim {\cal H}_\|$
, $\{{\cal H}_\|,{\cal H}_\perp\}\sim {\cal H}_\perp$ and $\{{\cal
H}_\perp,{\cal
H}_\perp \}\sim {\cal H}_\|$, where the crucial part is the last bracket. These
authors showed that in order to have path-independence of deformations, the
structure
function in the last bracket must equal the spatial metric on the hypersurface.
This
means that, if one knows the generator of parallel and normal deformations, it
is possible to read-off the spatial metric in the constraint algebra. (In a
canonical
field theory, the generators of local symmetries are represented by first class
constraints.) The parallel generator is singled out by the requirement that its
action
on the fundamental canonical fields should be that of the Lie-derivative. This
means
that here is :${\cal H}_\| =\tilde{\cal H}_a.$ And since the pair ${\cal H},
\tilde{\cal H}_a$ satisfies the needed algebra, I conclude that ${\cal H}_\perp
={\cal
H}$. That is because, if that would not be the case, the true generator of
normal
deformations could be written as: ${\cal H}_\perp = {\cal H} +
f^a(\Pi^{aI},A_{aI})\tilde{\cal H}_a$. Checking the constraint algebra for this
generator shows however that the required algebra is only found for
$f^a(\Pi^{aI},
A_{aI})=0$. So, here ${\cal H}_\|=\tilde{\cal H}_a$ and ${\cal H}_\perp = {\cal
H}$,
meaning that
the spatial metric can be read off in (\ref{3.16}). The spatial metric is
$q^{ab}=\Pi^{aI} \Pi^b_I$. Then, to find the time-time and time-space parts of
the
space-time metric, one should calculate the time-evolution of the spatial
metric.
According to Hojman, Kucha\v{r} and Teitelboim, it
is required that:

\begin{equation} \{q_{ab}(x),{\cal H}_{tot}\}=K_{ab}(x) \label{3.17}
\end{equation}

where $K_{ab}$ is the extrinsic curvature of the hypersurface embedded in
space-time.
Together with the normal definition of the extrinsic curvature in terms of the
space-time metric, this gives, with the help of the equations of motion, the
following
form of the space-time metric.

\begin{equation}
\tilde{g}^{\alpha \beta}=\sqrt{-g}g^{\alpha \beta}=\left(\begin{array}{cc}
-\frac{1}{N}& \frac{N^a}{N}\\
\frac{N^a}{N}& N q^{ab} -\frac{N^a N^b}{N}
\end{array} \right)\label{3.19}\end{equation}

That is, given any solution to the equations of motion following from
(\ref{3.11}),
the space-time metric is given by (\ref{3.19}). I do not know if this
definition of
 the metric,
which follows from purely geometrical considerations, always will fulfill the
other
requirements about geodesic, causal propagation of matter fields. Now,
following the
fields in (\ref{3.19}) backwards in the Legendre transform will give that the
object
called $g_{\alpha\beta}$ in section 2, really is the metric.

\subsection{Comparison to the Ashtekar and Witten Hamiltonians for gravity}
As shown in section 2, with the choice of $SO(1,2)$ as the gauge group, the
unified
theory describes ordinary Einstein gravity with a cosmological constant.
I will now show that the unified Hamiltonian, with gauge group $SO(1,2)$, is
related to the Witten \cite{Witt.2+1}, and to the Ashtekar \cite{Beng.2+1}
Hamiltonians via simple redefinitions of the constraints.

The Hamiltonian given by Witten, is

\begin{eqnarray}
{\cal H}^W_{tot}&=&N^I{\cal H}_I -A_{0I}{\cal G}^I \label{3.21} \\
 {\cal H}_I&:=&\Psi_I + \lambda f_{IJK}\Pi^{aJ}\Pi^{bK}\epsilon_{ab}\approx 0
\nonumber\\
{\cal G}^I&:=&{\cal D}_a\Pi^{aI}\approx 0 \nonumber \end{eqnarray}

which is just the Hamiltonian following from a Legendre transform from
(\ref{2.2}).
Now, it is easy to write the constraints in (\ref{3.11}) as linear combinations
of
those in (\ref{3.21}):

\begin{eqnarray}
{\cal H}&=&g^I(A_{aI},\Pi^{bJ}){\cal H}_I=\frac{1}{4\lambda} (\Psi^I-\lambda
f_{IJK}\Pi^{aJ}\Pi^{bK}\epsilon_{ab}){\cal H}_I \label{3.22}\\
{\cal H}_a&=&h^I_a(A_{aI},\Pi^{bJ}){\cal
H}_I=\frac{1}{2}\epsilon_{ab}\Pi^{bI}{\cal
H}_I \nonumber \end{eqnarray}

And, the Ashtekar Hamiltonian given by Bengtsson \cite{Beng.2+1}, is

\begin{eqnarray}
{\cal H}^{Ash}_{tot}&=&N{\cal H}^{Ash}+N^a{\cal H}^{Ash}_a -A_{0I}{\cal G}I
\label{3.23}\\
{\cal H}^{Ash}&:=&\frac{1}{4}\epsilon_{ab}\Pi^{aI}\Pi^{bJ}\Psi^Kf_{IJK} +
\frac{\lambda}{2} det(\Pi^{aI}\Pi^b_I)\approx 0 \nonumber \\
{\cal H}^{Ash}_a&:=&\frac{1}{2}\epsilon_{ab}\Pi^{bI}\Psi_I\approx 0 \nonumber
\end{eqnarray}

This Hamiltonian can be found by a Legendre transform from (\ref{2.3}), for
$SO(1,2)$.
It has the same form as the Ashtekar Hamiltonian for gravity in
(3+1)-dimensions
\cite{Ash}.
Again, the constraints in (\ref{3.11}) can be written as simple combinations of
the
constraints in (\ref{3.23}):

\begin{eqnarray}
{\cal H}&=&\Phi(A_{aI},\Pi^{bJ}){\cal H}^{Ash}=\frac{-1}{\lambda
det(\Pi^{aI}\Pi^b_I)}
(\frac{1}{4}
\epsilon_{ab}\Pi^{aI}\Pi^{bJ}\Psi^Kf_{IJK} - \frac{\lambda}{2}
det(\Pi^{aI}\Pi^b_I)){\cal H}^{Ash} \nonumber \\
{\cal H}_a&=&{\cal H}^{Ash}_a \label{3.24} \end{eqnarray}

This shows the equivalence between the three Hamiltonians (\ref{3.11}),
(\ref{3.21}) and (\ref{3.23}), for $SO(1,2)$ and non-degenerate metric.
Note however that, from (\ref{3.22}) and (\ref{3.24}) it is clear that the
unified
 Hamiltonian
(\ref{3.11}) is not totally equivalent to the Witten, and Ashtekar
Hamiltonians. That
is because the generalized Hamiltonian has solutions to the constraints
corresponding
to $g^I(A_{aI},\Pi^{bJ})=0$ or $\Phi(A_{aI},\Pi^{bJ})=0$. But this is a rather
innocent extension of the original Einstein theory, it just means that the
unified
theory has solutions corresponding to both signs of the cosmological constant
for a
given cosmological constant.
Classically this is of no importance while quantum mechanically this could lead
to
interesting effects, since the two different parts
of phase-space are connected via degenerate metric solutions.. What is more
important,
 is that
the unified theory really needs a cosmological constant. It can be arbitrarily
small, but it must be non-zero.

This all together shows the near-equivalence of the Hamiltonians (\ref{3.11}),
(\ref{3.21})
and (\ref{3.23}) for gauge group $SO(1,2)$ and $\lambda$ non-zero. What about
other
gauge groups?

For (\ref{3.21}) we still have a consistent theory for an arbitrary gauge
group, the
constraint algebra is closed. The constraint algebra will though not satisfy
the
general constraint algebra, given by Hojman, Kucha\v{r} and Teitelboim
\cite{HKT},
due to the fact that the constraints are not split in parallel and normal
generators of deformation. There is however another reason that makes it
less interesting
to generalize the Hamiltonian (\ref{3.21}) to other gauge groups.
This Hamiltonian will for an arbitrary gauge group always lack local
degrees of freedom. (The number of first class constraints are half the number
of
phase-space variables.)

How about (\ref{3.23}). This is the normal Ashtekar form of the constraints,
and it is
known that a crucial ingredients in the closure of the constraints algebra is
the
structure constants identity for $SO(1,2)$. For other gauge groups, which lack
this
identity, the constraint algebra will fail to close, and the theory will
produce
second class constraints.

So, it is only (\ref{3.11}) that can be generalized to arbitrary gauge groups
with a
closed constraint algebra and local degrees of freedom. The interesting aspect
of this
generalization is that it may be possible to find a new interacting theory of
gravity and Yang-Mills type of matter.

\subsection{Comparison to conventional Einstein-Yang-Mills theory.}
To compare the unified theory to the conventional Einstein-Yang-Mills theory, I
will choose a gauge group $G_{tot}=SO(1,2)\otimes G^{YM}$, and rewrite
the unified Hamiltonian (\ref{3.11}) in a form similar to the
conventional Hamiltonian. From there it is easy to show that the two
formulations
agree for weak Yang-Mills fields. Then, I will also show that the equations of
motion
for the Yang-Mills fields in the generalized theory is the normal Yang-Mills
equations
of motion, meaning that the changes come in the "Einstein's equations" part.

The conventional Hamiltonian for coupled Einstein-Yang-Mills is given by Romano
\cite{Romano}. See also Ashtekar, Romano and Tate \cite{ART}.

\begin{eqnarray}
{\cal H}^{conv}_{tot}&=&N{\cal H}^{conv}+N^a{\cal H}^{conv}_a - A_{0I}{\cal
G}^I
- A_{0i}{\cal G}^i \label{3.40} \\
{\cal H}^{conv}&:=&\frac{1}{4}\epsilon_{ab}\Pi^{aI}\Pi^{bJ}\Psi^Kf_{IJK} +
\frac{\lambda}{2}
det(\Pi^{aI}\Pi^b_I)+\frac{1}{4}(\epsilon_{ab}\epsilon_{cd}\Pi^{bI}\Pi^d_IE^{ai}E^c_i
+ B^iB_i) \nonumber\\
{\cal H}^{conv}_a&:=&\frac{1}{2}\epsilon_{ab}(\Pi^{bI}\Psi_I + E^{bi}B_i)
\nonumber\\
{\cal G}^{conv}_I&:=&{\cal D}_a\Pi^a_I \nonumber\\
{\cal G}_i&:=&{\cal D}_aE^a_i \nonumber \end{eqnarray}

where $\{A_{aI}(x),\Pi^{bJ}(y)\}=\delta^b_a\delta^J_I\delta^2(x-y)$ is the
conjugate
pair for the $SO(1,2)$ gravity part. And, $\{A_{ai}(x),E^{bj}(y)\}=\delta^b_a
\delta^j_i\delta^2(x-y)$ is the conjugate pair for the $G^{YM}$ Yang-Mills
matter
part. $B^i:=\epsilon^{ab}F^i_{ab}$, and the covariant derivative ${\cal D}_a$
"knows
how to act" on both $SO(1,2)$ and $G^{YM}$ indices. Note that in this
formulation, the
spatial metric is the square of $\Pi^{aI}$: $g^{ab}=\Pi^{aI}\Pi^b_I$

The unified theory, with gauge group $G_{tot}=SO(1,2)\otimes G^{YM}$, has the
Hamiltonian:

\begin{eqnarray}
{\cal H}_{tot}&=&N{\cal H}+N^a{\cal H}_a - A_{0I}{\cal G}^I
- A_{0i}{\cal G}^i \label{3.41}\\
{\cal H}&:=&\frac{1}{4\lambda}(\Psi^I\Psi_I + B^iB_i +
\lambda^2(det(\Pi^{aI}\Pi^b_I)
+\nonumber \\ &&+\epsilon_{ab}\epsilon_{cd}\Pi^{bI}\Pi^d_IE^{ai}E^c_i +
det(E^{ai}E^b_i))
 \nonumber\\
{\cal H}_a&:=&\frac{1}{2}\epsilon_{ab}(\Pi^{bI}\Psi_I + E^{bi}B_i) \nonumber\\
{\cal G}_I&:=&{\cal D}_a\Pi^a_I \nonumber\\
{\cal G}_i&:=&{\cal D}_aE^a_i \nonumber \end{eqnarray}

Remember that here is the spatial metric the "square" of the total momenta: $
g^{ab}=\Pi^{aI}\Pi^b_I+E^{ai}E^b_i$.
Now, to compare these two different formulations, I will rewrite ${\cal H}$ in
(\ref{3.41}) in a form similar to the form of ${\cal H}^{conv}$ in
(\ref{3.40}). To
do this, I first solve ${\cal H}_a=0$ for $\Psi^I$:

\begin{equation}
\Psi^I=\Phi(A_{aI},\Pi^{bJ},A_{ai},E^{bj})f^{IJK}\Pi^a_J\Pi^b_K\epsilon_{ab} -
\Pi^{aI} q_{ab} E^{bi}B_i \label{3.44} \end{equation}

where $\Phi(A_{aI},\Pi^{bJ},A_{ai},E^{bj})$ is a yet unknown function, and
$q_{ab}$ is
the inverse to $q^{ab}:=\Pi^{aI}\Pi^b_I$. Putting this into ${\cal H}=0$,
gives:

\begin{eqnarray}
\Phi(A_{aI},\Pi^{bJ},A_{ai},E^{bj})&=&\pm (\frac{\lambda^2}{4} +
\frac{1}{4det(q^{ab})}
(\lambda^2 \epsilon_{ab}\epsilon_{cd}\Pi^{bI}\Pi^d_IE^{ai}E^c_i + \lambda^2
det(E^{ai}E^b_i) + \nonumber \\
&&+ B^iB_i + B_iE^{ai}q_{ab}E^{bj}B_j))^{\frac{1}{2}} \label {3.45} \normalsize
\end{eqnarray}

Equations (\ref{3.44}) and (\ref{3.45}) give the total solution to the
diffeomorphism constraints ${\cal H}$ and ${\cal H}_a$. That means that
(\ref{3.44})
could be used together with (\ref{3.45}) as constraints instead of ${\cal H}$
and
 ${\cal H}_a$. I wanted however constraints that looked like the conventional
theory,
and to get that I simply multiply (\ref{3.44}) with
$f^{IJK}\Pi^a_J\Pi^b_k\epsilon_{ab}$. This finally gives the equivalent
constraints:

\begin{eqnarray} \small
&&\tilde{\cal H}=\frac{1}{4}\epsilon_{ab}\Pi^{aI}\Pi^{bJ}\Psi^Kf_{IJK} +
\nonumber \\
&&\pm\frac{|\lambda| det(q^{ab})}{2}\sqrt{1 + \frac{1}{det(q^{ab})}
 (\epsilon_{ab}\epsilon_{cd}\Pi^{bI}\Pi^d_IE^{ai}E^c_i + det(E^{ai}E^b_i)+
\frac{B^iB_i}{\lambda^2} +
\frac{B_iE^{ai}q_{ab}E^{bj}B_j}{\lambda^2})} \nonumber \\
&&{\cal H}_a:=\frac{1}{2}\epsilon_{ab}(\Pi^{bI}\Psi_I + E^{bi}B_i)\label{3.46}
\end{eqnarray} \normalsize

When comparing the conventional Hamiltonian (\ref{3.40}) and the unified
Hamiltonian (\ref{3.46}) it is important to note that $\Pi^{aI}$ in the two
different
theories does not have the same interpretation. In the conventional theory,
$\Pi^{aI}$
is the "square-root" of the metric, while in the generalized theory, the more
complicated relation $g^{ab}=\Pi^{aI}\Pi^b_I+E^{ai}E^b_i$ holds. However, for
weak
Yang-Mills fields, the two definitions coincide to lowest order in Yang-Mills
fields.
 Now, it is easy to do a series expansion of $\tilde{\cal H}$ for small
$E^{ai}$ and
$B^i$ around de-Sitter space. Remember that the unified theory equals
Einstein's
theory with a cosmological constant, when $G=SO(1,2)$, meaning that de-Sitter
space-time and vanishing Yang-Mills fields is a solution to the unified theory.
 So, $E^{ai}E^b_i\ll 1, \hspace{5mm}B^iB_i\ll \lambda^2$
and $\Pi^{aI}\Pi^b_I \sim 1 \Rightarrow$

\begin{eqnarray}
\tilde{\cal H}&=&\frac{1}{4}\epsilon_{ab}\Pi^{aI}\Pi^{bJ}\Psi^Kf_{IJK} \pm
\frac{|\lambda|}{2} det(\Pi^{aI}\Pi^b_I)\label{3.47} \\
&&\pm  \frac{|\lambda|}{4}(\epsilon_{ab}\epsilon_{cd}\Pi^{bI}\Pi^d_IE^{ai}E^c_i
+ \frac{B^iB_i}{\lambda^2}) +
\vartheta((E^{ai}E^b_i)^2,\frac{(B^iB_i)^2}{\lambda^4})
\nonumber\end{eqnarray}

which shows that the unified theory agrees with the conventional theory to
lowest
order in weak Yang-Mills fields on approximately de-Sitter space-time, if one
rescales the Yang-Mills fields: $\tilde{E}^{ai}:=\sqrt{|\lambda|}E^{ai},
\hspace{5mm}
\tilde{B}^i:=\frac{1}{\sqrt{|\lambda|}} B^i$. $\tilde{E}^{ai}$ and
$\tilde{B}^i$ is to
be interpreted as the physical Yang-Mills fields. The $\pm$ signs in front of
the
Yang-Mills coupling can always be absorbed in the $G^{YM}$ "group-metric".

The question is then: What kind of corrections to the normal theory do we get
for
strong Yang-Mills fields? I will now show that it is only the "Einstein's
equation"
part of the theory that changes. The Yang-Mills equation of motion remains
unaltered.
 This
is easily seen by comparing the equation of motion for the Yang-Mills fields
coming
from (\ref{3.40}) and (\ref{3.41}).

Conventional theory:

\begin{eqnarray}
&&\dot{A}_{ai}-{\cal
D}_aA_{0I}-\frac{N}{2}\Pi^{cI}\Pi^d_I\epsilon_{ca}\epsilon_{db}
E^b_i -\frac{1}{2} N^b\epsilon_{ba}B_i=0 \label{3.50}\\
&&-\dot{E}^{ai}+f^{ijk}E^a_jA_{0k} +\epsilon^{ca}{\cal D}_c(N B^i) +
\epsilon_{ca}{\cal
D}_c(N^e\epsilon_{eb}E^{bi})=0 \label{3.51} \end{eqnarray}

And the unified theory:

\begin{eqnarray}
&&\dot{A}_{ai}-{\cal D}_aA_{0I}-\frac{N\lambda}{2}(\Pi^{cI}\Pi^d_I+
E^{cj}E^d_j)
\epsilon_{ca}\epsilon_{db}E^b_i-\frac{1}{2} N^b\epsilon_{ba}B_i=0
\label{3.52}\\
&&-\dot{E}^{ai}+f^{ijk}E^a_jA_{0k} +\epsilon^{ca}{\cal
D}_c(\frac{N}{\lambda}B^i)+
\epsilon^{ca}{\cal D}_c(N^e\epsilon_{eb}E^{bi})=0 \label{3.53}
\end{eqnarray}

which together with the fact that $q^{cd}=\Pi^{cI}\Pi^d_I$ in the conventional
theory,
and $q^{cd}=\Pi^{cI}\Pi^d_I+ E^{cj}E^d_j$ in the generalized theory, gives that
the
physical Yang-Mills fields should satisfy the same equation of motion in both
theories. The equation of motion for the gravity part will however be different
in
the two theories. This will become even clearer in the second order
formulation, in
the next section.
\section{Second order, pure connection Lagrangian formulation.}
In this section, I will derive a second order pure connection Lagrangian
corresponding
to the unified theory (\ref{2.1}) and (\ref{3.11}). This second order
formulation
is most easily found by eliminating the $e_{\alpha I}$ field from the first
order
Lagrangian (\ref{2.1}). It can also be found by performing a Legendre transform
from
the Hamiltonian (\ref{3.11}), treating the diffeomorphism constraints as
primary
constraints. I will use the former method here.

I start with the first order Lagrangian:

\begin{equation} {\cal L}= \epsilon^{\alpha \beta \gamma} e_{\alpha I}
 F^I_{\beta \gamma}
+ \lambda \sqrt{-g} \label{4.1} \end{equation}

Now, I want to eliminate the $e_{\alpha I}$ field from the Lagrangian, by
solving its
equation of motion.

\begin{equation}
\frac{\delta S}{\delta e_{\alpha I}}=F^{\ast\alpha I}+\lambda\sqrt{-g}e^{\alpha
I}=0
\label{4.2} \end{equation}

where $F^{\ast\alpha I}:=\epsilon^{\alpha\beta\gamma}F^I_{\beta\gamma}$.
Remember
that\\
$g_{\alpha\beta}:=e_{\alpha I}e_\beta^I,
\hspace{5mm}g^{\alpha\beta}g_{\beta\gamma}
=\delta^\alpha_\gamma, \hspace{5mm}g=\frac{1}{6}\epsilon^{\alpha \beta \gamma}
\epsilon^{\delta \epsilon \sigma} g_{\alpha \delta} g_{\beta \epsilon}
g_{\gamma \sigma}, \hspace{5mm}e^{\alpha I}:=g^{\alpha\beta}e_\beta^I$.
Equation
(\ref{4.2}) gives that

\begin{equation}
F^{\ast\alpha I}F^{\ast\beta}_I=-\lambda^2gg^{\alpha\beta} \label{4.3}
\end{equation}

Taking the determinant of both sides, yields

\begin{equation}
det(F^{\ast\alpha I}F^{\ast\beta}_I)=-\lambda^6g^2 \label{4.4} \end{equation}

Now, (\ref{4.2}) and (\ref{4.4}) give

\begin{equation}
e^{\alpha I}=sign(-\lambda)(-\frac{1}{\lambda^2}det(F^{\ast\alpha I}
F^{\ast\beta}_I))^{-\frac{1}{4}} F^{\ast\alpha I} \label{4.5} \end{equation}

Putting this into (\ref{4.1}), gives

\begin{equation}
{\cal L}=-2
sign(\lambda)(-\frac{1}{\lambda^2}det(F^{\ast\alpha
I}F^{\ast\beta}_I))^\frac{1}{4}
\label{4.6} \end{equation}

which is the wanted pure connection, second order Lagrangian for the unified
theory. For gauge group $SO(1,2)$ it is possible to simplify the Lagrangian
slightly
by using the relation: $det(F^{\ast\alpha
I}F^{\ast\beta}_I)=-(det(F^{\ast\alpha I}))^2,
 \hspace{5mm}det(F^{\ast\alpha I})=
\frac{1}{6}\epsilon_{\alpha \beta \gamma}f_{IJK} F^{\ast\alpha I} F^{\ast\beta
J}
F^{\ast\gamma K}$. This gives the Lagrangian

\begin{equation}
{\cal L}^{SO(1,2)}=-\frac{2}{\lambda}\sqrt{det(F^{\ast\alpha I})} \label{4.65}
\end{equation}

which is the Lagrangian that was found in \cite{PP2+1}, using a Legendre
transform
from Ashtekar's Hamiltonian for gravity with a cosmological constant.
\subsection{Metric formulas.}
To find the metric formulas in this pure connection formulation, I start with
the
expression for the space-time metric (\ref{3.19}) that followed from Hojman,
Kucha\v{r} and Teitelboim's \cite{HKT} "metric definition". Then one can just
follow
the fields through the Legendre transform. The metric formula is

\begin{equation}
\tilde{g}^{\alpha \beta}=\sqrt{-g}g^{\alpha \beta}=\left(\begin{array}{cc}
-\frac{1}{N}& \frac{N^a}{N}\\
\frac{N^a}{N}& N \Pi^{aI}\Pi^b_I -\frac{N^a N^b}{N}
\end{array} \right)\label{4.7} \end{equation}

given in terms of the phase-space coordinates and Lagrange multiplier fields.
Now,
performing the Legendre transform, and following the fields through, gives:

\begin{eqnarray}
\tilde{g}^{\alpha\beta}=\sqrt{-g}g^{\alpha\beta}&=&\frac{1}{\lambda^2\sqrt{-g}}
F^{\ast\alpha I} F^{\ast\beta}_I \label{4.8}\\
\sqrt{-g}&=&\frac{1}{|\lambda|}(-\frac{1}{\lambda^2}
det(F^{\ast\alpha I}F^{\ast\beta}_I))^\frac{1}{4} \label{4.9}\\
g^{\alpha\beta}&=&(-\frac{1}{\lambda^2}
det(F^{\ast\alpha I}F^{\ast\beta}_I))^{-\frac{1}{2}} F^{\ast\alpha I}
F^{\ast\beta}_I
\label{4.10}\\
g_{\alpha\beta}&=&-\frac{2}{\lambda^2} ((-\frac{1}{\lambda^2}
det(F^{\ast\alpha I}F^{\ast\beta}_I))^{-\frac{1}{2}}
F^J_{\alpha\gamma}F^{\ast\gamma I}
F^{\ast\sigma}_I F_{\sigma\beta J} \label{4.11} \end{eqnarray}

\subsection{Equations of motion}
Finally in this section, I will show that the equations of motion following
from the
Lagrangian (\ref{4.6}) is just the ordinary Yang-Mills equations. Varying the
Lagrangian (\ref{4.6}) with respect to the connection $A_{\alpha I}$, and using
the
metric-formulas (\ref{4.8})-(\ref{4.11}) gives

\begin{equation}
\frac{\delta S}{\delta A_{\alpha I}}={\cal D}_\beta(\sqrt{-g}g^{\beta\gamma}
g^{\alpha\sigma} F^I_{\gamma\sigma})=0 \label{4.12} \end{equation}

the Yang-Mills equations of motion. Note however that here is the metric-field
not a
non-dynamically background field. The metric is a function of the connection.
Since
this Lagrangian also is valid for pure gravity with a cosmological constant
this means
that {\em Einstein gravity with a cosmological constant in (2+1)-dimensions
really is
an $SO(1,2)$ Yang-Mills theory, on its own dynamical background space-time}!
For other gauge
groups, like $G^{tot}=SO(1,2)\otimes G^{YM}$, this means that the equation of
motion
for the Yang-Mills field is the normal one, while Einstein's equation is
changed.

\section{Static, rotation symmetric solutions to Einstein-Maxwell theory.}
To compare the unified theory, presented in section 2-4, with the conventional
Einstein-Yang-Mills theory, I will here study some simple explicit solutions to
both
theories. I will study the Einstein-Maxwell theory, which means choosing gauge
group
$G^{tot}=SO(1,2)\otimes U(1)$ for the unified theory. To simplify the theories
as
much as possible, I will only look for static and rotation symmetric solutions.
These
kind of solutions has previously been studied for Einstein-Maxwell without a
cosmological constant, by Deser and Mazur \cite{Deser} and Melvin
\cite{Melvin}.
However, for a non-vanishing cosmological constant, I have not been able to
find any
work done on explicit solutions to this theory, so I will start by deriving
solutions
to the conventional theory. Then, the unified theory will be solved under the
same
assumptions of symmetries, and the explicit solutions will be compared. In the
solutions it will be clear that the solution to the conventional theory is the
lowest
order term, in Maxwell fields, in the solution to the unified theory.
Otherwise,
the only interesting new feature for the solution to the generalized theory, is
that,
for a static electric charge, there exist no solution to the unified theory,
inside a circle of radius $r=q$ in Schwardschild coordinates.
\subsection{Conventional theory.}
The conventional Einstein-Maxwell theory with a cosmological constant was given
in
(\ref{3.40}).

\begin{eqnarray}
{\cal H}^{conv}_{tot}&=&N{\cal H}^{conv}+N^a{\cal H}^{conv}_a - A_{0I}{\cal
G}^I
- A_{0}{\cal G} \label{5.1} \\
{\cal H}^{conv}&:=&\frac{1}{4}\epsilon_{ab}\Pi^{aI}\Pi^{bJ}\Psi^Kf_{IJK} +
\frac{\lambda}{2}
det(\Pi^{aI}\Pi^b_I)+\frac{1}{4}(\epsilon_{ab}\epsilon_{cd}\Pi^{bI}\Pi^d_IE^{a}E^c
+ B^2)\approx 0 \nonumber\\
{\cal H}^{conv}_a&:=&\frac{1}{2}\epsilon_{ab}(\Pi^{bI}\Psi_I + E^{b}B)\approx 0
 \nonumber\\
{\cal G}^{conv}_I&:=&{\cal D}_a\Pi^a_I\approx 0 \nonumber\\
{\cal G}&:=&\partial _a E^a\approx 0 \nonumber \end{eqnarray}

and the space-time metric is

\begin{equation}
\tilde{g}^{\alpha \beta}=\sqrt{-g}g^{\alpha \beta}=\left(\begin{array}{cc}
-\frac{1}{N}& \frac{N^a}{N}\\
\frac{N^a}{N}& N \Pi^{aI}\Pi^b_I -\frac{N^a N^b}{N}
\end{array} \right)\label{5.2}\end{equation}

Now, a static and rotation symmetric metric can always be put in a form

\begin{equation}
g_{\alpha \beta}=\left(\begin{array}{ccc}
-\xi^2(r)&0&0\\
0&\chi^2(r)&0\\
0&0&\psi^2(r)\end{array} \right)\label{5.3}\end{equation}

where $\xi (r), \chi (r)$ and $\psi (r)$ are three arbitrary functions of some
radial
coordinate $r$. One could also fix the $r$-coordinate gauge by e.g choosing
$\psi
(r)=r$, which corresponds to Schwardschild coordinates, but I will not fix that
gauge
yet. This static, rotation symmetric metric then means

\begin{equation}
N^a=0, \hspace{10mm}\Pi^{rI}\Pi^\theta_I=0 \label{5.4} \end{equation}

But I still have to fix the $SO(1,2)$ and $U(1)$ gauge. The $U(1)$ gauge should
be
fixed so that the gauge choice one makes always can be reached by a $U(1)$
gauge
transformation from an arbitrary static and rotation symmetric field
configuration.
This singles out $A_r=0$ as a good gauge choice. For the $SO(1,2)$ gauge, one
has the
same requirements plus that one wants the spatial metric to be positive
definite.
Therefore, I make the choices: $\Pi^{rI}=(0,\psi(r),0), \Pi^{\theta
I}=(0,0,\chi(r))$.
This is three $SO(1,2)$ gauge choices together with the orthogonality condition
(\ref{5.4}). Note that this gauge choice will agree with the static and
rotation
symmetric metric (\ref{5.3}), if $N=\frac{\xi(r)}{\chi(r) \psi(r)}$. Now, it is
a
straightforward task to put in this static and rotation symmetric Ansatz
together
with the gauge choices, into the constraints and equations of motion following
from
the Hamiltonian (\ref{5.1}). This will rather immediately give that

\begin{equation}
E^r=q,\hspace{5mm}E^\theta=0,\hspace{5mm}B=\frac{k}{N(r)},\hspace{5mm}qk=0
\label{5.45}
\end{equation}
where $q$ and $k$ are konstants. This means that one cannot have both electric
and
magnetic static, rotation symmetric fields in the same solution. I start with
the
electric solution.
\subsubsection{Electric solution to conventional theory.}
Electric solution means $k=0$ and $q\neq0$. Using this in the equations of
motion and
constraints gives

\begin{equation}\begin{array}{lll}
E^r=q& E^\theta=0& B=0\\
A_{01}=0& A_{03}=0\\
A_{\theta 2}=0& A_{\theta 3}=0\\
A_{r1}=0& A_{r2}=0& A_{r3}=0 \end{array} \label{5.5} \end{equation}

and defining $\xi(r):=N(r)\psi (r) \chi (r)$ gives the equations:

\begin{eqnarray}
A^\prime_0&=&-\frac{q}{2}\frac{\xi (r)\chi (r)}{\psi (r)} \label{5.55} \\
A_{02}&=&\frac{\xi^\prime (r)}{\chi (r)} \label{5.6} \\
A^\prime_{02}&=&-\frac{\lambda}{2}\xi (r) \chi (r) + \frac{q^2}{4} \frac{\xi
(r) \chi
(r)}{\psi^2(r)} \label{5.7}\\
A^1_{\theta}A_{02}&=&\frac{\lambda}{2}\xi (r) \psi (r) + \frac{q^2}{4}
\frac{\xi
(r)}{\psi (r)} \label{5.8}\\
A^1_{\theta}&=&-\frac{\psi^\prime (r)}{\chi (r)} \label{5.9}\\
A^{\prime 1}_\theta&=&\frac{\lambda}{2}\psi (r)\chi (r)+\frac{q^2}{4}\frac{\chi
(r)}{\psi (r)} \label{5.10} \end{eqnarray}

where the prime denotes differentiation with respect to $r$. Remember that the
$SO(1,2)$ "group-metric" is $\eta_{IJ}=diag(-1,1,1)$, meaning that
$A_0^1=-A_{01}$.
Now it is time to fix the $r$-coordinate gauge, and this is done here by
choosing
Schwardschild coordinates: $g_{\theta \theta}=\psi^2(r)=r^2$. Using this gauge,
it is
an easy task to solve all the equations (\ref{5.6})-(\ref{5.10}), yielding

\begin{eqnarray}
\chi (r)&=&\frac{1}{\sqrt{C_1-\frac{\lambda r^2}{2} -\frac{q^2}{2} log(r)}}
\label{5.11}\\
\xi (r)&=&C_2 \sqrt{C_1-\frac{\lambda r^2}{2} -\frac{q^2}{2} log(r)}
\label{5.12}
\end{eqnarray}

where $C_1$ and $C_2$ are constants of integration. This gives for the metric

\begin{equation} g_{\alpha \beta}= \left( \begin{array}{ccc}
-(C_2)^2(C_1-\frac{\lambda r^2}{2} -\frac{q^2}{2} log(r))&0&\\
0&(C_1-\frac{\lambda r^2}{2} -\frac{q^2}{2} log(r))^{-1}&0\\
0&0&r^2 \end{array}\right) \label{5.13} \end{equation}

If one puts $\lambda=0$, one recovers the solution found by Deser and Mazur
\cite{Deser}
and Melvin \cite{Melvin}. The solution for the connection
 follows from equations (\ref{5.55})-(\ref{5.10}).
If one want to relate this solution to the normal covariant Einstein equation,
the
solution (\ref{5.13}) solves the equations:
\begin{eqnarray}
&&R_{\alpha\beta}-\frac{1}{2}Rg_{\alpha\beta}=\frac{\lambda}{2}g_{\alpha\beta}-
T_{\alpha\beta} \label{5.145} \\
&&T_{\alpha\beta}=2F_{\alpha\gamma}g^{\gamma\sigma}F_{\beta\sigma}-\frac{1}{2}g_{\alpha
\beta}(g^{\gamma\sigma}g^{\epsilon\delta}F_{\gamma\epsilon}F_{\sigma\delta})
\nonumber\\
&&\partial_\alpha(\sqrt{-g}g^{\alpha\beta}g^{\gamma\sigma}F_{\beta\sigma})=0
\nonumber
\end{eqnarray}

with

\begin{equation} F_{\alpha\beta}=\left(\begin{array}{ccc}
0&-\frac{C_2q}{r}&0\\
\frac{C_2q}{r}&0&0\\
0&0&0 \end{array}\right) \label{5.155} \end{equation}

The normal factor, $8\pi G$ is included in $q$.

\subsubsection{Magnetic solution to conventional theory.}
Magnetic solution means that $q=0$ but $k\neq 0$. This gives
\begin{equation} \begin{array}{lll}
E^r=0&E^\theta=0&B(r)=\frac{k}{N(r)}\\
A_{01}=0& A_{03}=0\\
A_{\theta 2}=0& A_{\theta 3}=0\\
A_{r1}=0& A_{r2}=0& A_{r3}=0 \end{array} \label{5.14} \end{equation}

and again defining $\xi(r):=N(r)\psi (r) \chi (r)$ gives the equations:

\begin{eqnarray}
A_0&=&constant \label{5.15}\\
A_{02}&=&\frac{\xi^\prime (r)}{\chi (r)} \label{5.16}\\
A^\prime_{02}&=&-\frac{\lambda}{2}\xi (r) \chi (r) + \frac{k^2}{4} \frac{\chi
(r)}
{\xi (r)} \label{5.17}\\
A^1_{\theta}A_{02}&=&\frac{\lambda}{2}\xi (r) \psi (r) - \frac{k^2}{4}
\frac{\psi
(r)}{\xi (r)} \label{5.18}\\
A^1_{\theta}&=&-\frac{\psi^\prime (r)}{\chi (r)} \label{5.19}\\
A^{\prime 1}_\theta&=&\frac{\lambda}{2}\psi (r)\chi (r)+\frac{k^2}{4}\frac{\chi
(r)\psi (r)}{\xi^2 (r)} \label{5.20} \end{eqnarray}

 Now it is time to fix the $r$-coordinate
gauge, and the reason why I did not choose the Schwardschild gauge from the
beginning,
is that in choosing that gauge, I have not been able to find an explicit
solution for
the metric. That is, with the gauge choice $g_{\theta\theta}=\psi^2 (r)=r^2$,
the
equations (\ref{5.16})-(\ref{5.20}) has no simple explicit solution for $\xi
(r)$ and
$\chi (r)$.(One gets relations like:$-\frac{\lambda}{4} \xi^2 (r)
+\frac{k^2}{4}
log\xi (r)=D r$) So instead I try the gauge choice $\xi (r)=r$. with this gauge
choice it is easy to find the solution to (\ref{5.16})-(\ref{5.20}).

\begin{eqnarray}
\chi (r)&=&\frac{1}{\sqrt{C_3-\frac{\lambda r^2}{2} +\frac{k^2}{2} log(r)}}
\label{5.21}\\
\psi (r)&=&C_4 \sqrt{C_3-\frac{\lambda r^2}{2} +\frac{k^2}{2} log(r)}
\label{5.22}
\end{eqnarray}

where $C_3$ and $C_4$ are constants of integration. This gives for the metric

\begin{equation} g_{\alpha \beta}= \left( \begin{array}{ccc}
-r^2&0&\\
0&(C_3-\frac{\lambda r^2}{2} +\frac{k^2}{2} log(r))^{-1}&0\\
0&0&(C_4)^2(C_3-\frac{\lambda r^2}{2} +\frac{k^2}{2} log(r)) \end{array}\right)
\label{5.23} \end{equation}

which shows a remarkable symmetry between the electric and magnetic solution.
If one
puts $\lambda=0$ it is possible to change coordinates into Schwardschild or
conformal
coordinates, and this solution will agree with Melvin's solution \cite{Melvin}.

In terms of the covariant Einstein equation, this solution solves
(\ref{5.145}), with

\begin{equation} F_{\alpha\beta}=\left(\begin{array}{ccc}
0&0&0\\
0&0&\frac{C_4k}{r}\\
0&-\frac{C_4k}{r}&0 \end{array}\right) \label{5.255} \end{equation}

\subsection{The unified theory.}
The unified theory with gauge group $G^{tot}=SO(1,2)\otimes U(1)$ is described
by

\begin{eqnarray}
{\cal H}_{tot}&=&N{\cal H}+N^a{\cal H}_a - A_{0I}{\cal G}^I
- A_{0}{\cal G} \label{6.1}\\
{\cal H}&:=&\frac{1}{4\lambda}(\Psi^I\Psi_I + B^2 +
\lambda^2(det(\Pi^{aI}\Pi^b_I)
+\nonumber \\ &&+\epsilon_{ab}\epsilon_{cd}\Pi^{bI}\Pi^d_IE^{a}E^c)\approx 0
 \nonumber\\
{\cal H}_a&:=&\frac{1}{2}\epsilon_{ab}(\Pi^{bI}\Psi_I + E^{b}B)\approx 0
\nonumber\\
{\cal G}_I&:=&{\cal D}_a\Pi^a_I\approx 0 \nonumber\\
{\cal G}&:=&\partial_aE^a\approx 0 \nonumber \end{eqnarray}

where I have chosen the total "group-metric" to be $\eta_{ij}=diag(-1,1,1,1)$
and
$i=I$ for $1\leq i\leq 3$ and $i=4$ concerns the $U(1)$ fields. Note that there
is
nothing unique about this choice of "group-metric", I could equally well have
chosen
$\eta_{44}=-1$. The only requirement the "group-metric" has to fulfill is that
it
should be left invariant under gauge transformations. The densitized space-time
metric
is given by the formula

\begin{equation}
\tilde{g}^{\alpha \beta}=\sqrt{-g}g^{\alpha \beta}=\left(\begin{array}{cc}
-\frac{1}{N}& \frac{N^a}{N}\\
\frac{N^a}{N}& N g^{ab} -\frac{N^a N^b}{N}
\end{array} \right)\label{6.2}\end{equation}

where $g^{ab}=\Pi^{aI}\Pi^b_I+E^aE^b$ now. The static and rotation symmetric
Ansatz is
the same as for the conventional theory:

\begin{equation}
g_{\alpha \beta}=\left(\begin{array}{ccc}
-\xi^2(r)&0&0\\
0&\chi^2(r)&0\\
0&0&\psi^2(r)\end{array} \right)\label{6.3}\end{equation}

where $\xi (r), \chi (r)$ and $\psi (r)$ are three arbitrary functions of some
radial
coordinate $r$. The only coordinate-gauge that is left to fix is the
$r$-coordinate.
This Ansatz then means

\begin{equation}
N^a=0, \hspace{10mm}\Pi^{rI}\Pi^\theta_I=0 \label{6.4} \end{equation}

Then it is time to fix the $SO(1,2)$ and $U(1)$ gauge. The $U(1)$ gauge is
fixed in
the same way as for the conventional theory: $A_r=0$. But, in fixing the
$SO(1,2)$
gauge one must now be more careful. What is required is again that the gauge
choice
could be reached by a $SO(1,2)$ gauge transformation from an arbitrary static
and
rotation symmetric field configuration. And that the spatial metric $g^{ab}$
should be
positive definite. But since $g^{ab}=\Pi^{aI}\Pi^b_I+E^aE^b$ this means that
one must
allow $\Pi^{aI}$ to be time-like with respect to the $SO(1,2)$ "group-metric".
Since
the solution to ${\cal G}=0$ and $\dot{A}_\theta-\{A_\theta,H\}=0$ is $E^r=q$
and
$E^\theta=0$, a good gauge choice respecting the mentioned requirements should
be:
$E^{rI}=(q,\gamma (r),0)$ and $E^{\theta I}=(0,0,\chi (r))$. For $\gamma (r) >
q$ it
is always possible to gauge-rotate (boost) $E^{rI}$ into
$\tilde{E}^{rI}=(0,\psi
(r)=\sqrt{\gamma^2 (r)-q^2},0)$, and for $\gamma (r) < q$ it
is always possible to gauge-rotate (boost) $E^{rI}$ into
$\tilde{\tilde{E}}^{rI}=(\sqrt{q^2-\gamma^2 (r)},0,0)$. One can then go on and
solve
the constraints and equations of motion in the two different regions $\gamma
(r) >
q$ and $\gamma (r) < q$, and then glue the solutions back together. In doing
this,
one soon notice that there exist no real solution to the constraints ${\cal
H}=0$
and ${\cal H}_a=0$ for $\gamma (r)$ in the region $\gamma (r)\leq q$. This
means that
it is in fact okey to use the same gauge choice here as in the conventional
theory:
$E^{rI}=(0,\psi (r),0)$ and $E^{\theta I}=(0,0,\chi (r))$.

Now, using these gauge choices together with the static and rotation symmetric
Ansatz
in the constraints and equations of motion following from (\ref{6.1}), gives

\begin{equation}
E^r=q,\hspace{5mm}E^\theta=0,\hspace{5mm}B=\frac{k}{N(r)},\hspace{5mm}qk=0
\label{6.45}
\end{equation}
where $q$ and $k$ are konstants. This again means that one cannot have both
electric
and
magnetic static, rotation symmetric fields in the same solution. I start with
the
electric solution.
\subsubsection{Electric solution to the unified theory.}
Electric solution: $q\neq 0$ and $B=0$. Using this in the equations, gives

\begin{equation}\begin{array}{lll}
E^r=q& E^\theta=0& B=0\\
A_{01}=0& A_{03}=0\\
A_{\theta 2}=0& A_{\theta 3}=0\\
A_{r1}=0& A_{r2}=0& A_{r3}=0 \end{array} \label{6.5} \end{equation}

and defining $\xi(r):=N(r)\psi (r) \chi (r)$ gives the equations:

\begin{eqnarray}
A^\prime_0&=&-\frac{\lambda q}{2}\frac{\xi (r)\chi (r)}{\psi (r)} \label{6.55}
\\
A_{02}&=&\pm sign(\lambda)\frac{\xi^\prime (r)}{\chi (r)} \label{6.6} \\
A^\prime_{02}&=&- \frac{\lambda}{2}\frac{\sqrt{\psi^2(r)-q^2}}{\psi (r)}
\xi (r) \chi (r)  \label{6.7}\\
A^1_{\theta}A_{02}&=&\frac{\lambda}{2}\xi (r) \psi (r)  \label{6.8}\\
A^1_{\theta}&=&-\frac{\psi (r)}{\sqrt{\psi^2 (r)-q^2}}
\frac{\psi^\prime (r)}{\chi (r)} \label{6.9}\\
A^{\prime 1}_\theta&=&\pm \frac{|\lambda|}{2}\psi (r)\chi (r) \label{6.10}
\end{eqnarray}

Now, choosing the Schwardschild gauge $g_{\theta\theta}=\psi^2(r)=r^2$ gives
the
solution

\begin{eqnarray}
\chi (r)&=&\frac{r}{\sqrt{r^2-q^2}}\frac{1}{\sqrt{D_1\mp \frac{|\lambda|r}{2}
\sqrt{r^2-q^2} \mp \frac{|\lambda|q^2}{2}log(r+\sqrt{r^2-q^2})}} \label{6.11}
\\
\xi (r)&=& D_2\sqrt{D_1\mp \frac{|\lambda|r}{2}
\sqrt{r^2-q^2} \mp \frac{|\lambda|q^2}{2}log(r+\sqrt{r^2-q^2})} \label{6.12}
\end{eqnarray}

where $D_1$ and $D_2$ are constants of integration. This gives for the
space-time
metric

\begin{eqnarray}
g_{tt}&=&-(D_2)^2(D_1\mp \frac{|\lambda|r}{2}
\sqrt{r^2-q^2} \mp \frac{|\lambda|q^2}{2}log(r+\sqrt{r^2-q^2})) \label{6.13} \\
g_{rr}&=&\frac{r^2}{r^2-q^2}(D_1\mp \frac{|\lambda|r}{2}
\sqrt{r^2-q^2} \mp \frac{|\lambda|q^2}{2}log(r+\sqrt{r^2-q^2}))^{-1} \nonumber
\\
g_{\theta\theta}&=&r^2 \nonumber \end{eqnarray}

Doing a Taylor series expansion of the metric for $q\ll r$ gives

\begin{eqnarray}
g_{tt}&=&-(D_2)^2(\tilde{D}_1\mp \frac{|\lambda|r^2}{2}\mp
\frac{|\lambda|q^2}{2}log(r))+\vartheta (\frac{q^2}{r^2}) \label{6.14} \\
g_{rr}&=&(\tilde{D}_1\mp \frac{|\lambda|r^2}{2}\mp
\frac{|\lambda|q^2}{2}log(r))^{-1}+\vartheta (\frac{q^2}{r^2}) \nonumber\\
g_{\theta\theta}&=&r^2 \nonumber \end{eqnarray}

where $\tilde{D}_1=D_1\pm
\frac{|\lambda|q^2}{4}\mp\frac{|\lambda|q^2log(2)}{2}$.
This shows that the solution to the generalized theory agrees with the solution
to the
conventional theory, to lowest order in the Maxwell fields. Note that the
physical
electric field is the rescaled $E$: $E^a_{phys}=\sqrt{|\lambda |}E^a$.

Here in the explicit solution (\ref{6.13}) it is clear that this solution is
only
valid outside a circle of radius $r=q$, and this is not just a coordinate
singularity.
Calculating the invariant curvature scalar $R:=g^{\alpha\beta}R_{\alpha\beta}$,
gives
that $R$ becomes complex inside this radius. It is also clear from the form of
the
constraints ${\cal H}$ and ${\cal H}_a$ that with $B=0$ and $\Pi^{aI}$
time-like, there
exist no real solution for $\Psi^I$. (Remember that $q_{\theta\theta}\leq q$
implies
that $\Pi^{rI}$ is time-like.) The conclusion of this must then be that
space-time
does not exists inside a radius $r=q$ around a point-charge. Or, electrical
point-charges is not allowed, charged objects must have extensions. If this
result
would be valid in (3+1)-dimensions, we could calculate what distance $r=q$
corresponds
to for an electron: $r\sim \frac{e \sqrt{G}}{\sqrt{\epsilon_0} c^2}\sim
10^{-26}m$.
However, if this result is a purely (2+1)-dimensional effect, we do not know
the
constants of nature, and therefore cannot do this calculation.

What kind of "singularity" is $r=q$ then? Does the curvature invariants
diverge, or is
the invariant distance to this limit infinite? I have only been able to check
one
curvature invariant, R, and that invariant is finite and well behaved in the
limit
$r\rightarrow q$

\begin{equation}
R\sim \frac{2C_1}{q^2} -\lambda log(q) + \vartheta(\sqrt{r-q}) \label{6.145}
\end{equation}

It is also clear that the invariant length $\int^q_{r_0}\sqrt{g_{rr}} dr$ must
be finite
since $g_{rr}\sim \frac{1}{r-q}$ for $r\sim q$, which means that the integral
is
finite.

\subsubsection{Magnetic solution to the unified theory.}
Magnetic solution means: $q=0$ and $B\neq 0$. Using this together with the
static and
rotation symmetric Ansatz as well as the gauge choices, gives

\begin{equation}\begin{array}{lll}
E^r=0& E^\theta=0& B(r)=\frac{k}{N(r)}\\
A_{01}=0& A_{03}=0\\
A_{\theta 2}=0& A_{\theta 3}=0\\
A_{r1}=0& A_{r2}=0& A_{r3}=0 \end{array} \label{7.5} \end{equation}

and defining $\xi(r):=N(r)\psi (r) \chi (r)$ gives the equations:

\begin{eqnarray}
A_0&=&konst \label{7.55} \\
A_{02}&=&\pm sign(\lambda) \frac{ \partial _r \sqrt{\xi^2(r)+
\frac{k^2}{\lambda^2} }}
{\chi (r)} \label{7.6} \\
A^\prime_{02}&=&- \frac{\lambda}{2} \xi (r) \chi (r)  \label{7.7}\\
A^1_{\theta}A_{02}&=&\frac{\lambda}{2}\xi (r) \psi (r)  \label{7.8}\\
A^1_{\theta}&=&-\frac{\psi^\prime (r)}{\chi (r)} \label{7.9}\\
A^{\prime 1}_\theta&=&\pm \frac{|\lambda|}{2}
\frac{\sqrt{\xi^2(r)+\frac{k^2}{\lambda^2}}}{\xi (r)}\psi (r)\chi (r)
\label{7.10}
\end{eqnarray}

Here I fix the $r$-coordinate gauge like I did for the conventional theory:
$\xi
(r)=r$. This gives the solution

\begin{eqnarray}
\chi (r)&=&\frac{r}{\sqrt{r^2+\frac{k^2}{\lambda^2}}}
\frac{1}{\sqrt{D_3\mp \frac{|\lambda|r}{2}
\sqrt{r^2+\frac{k^2}{\lambda^2}} \pm
\frac{|\lambda|k^2}{2\lambda^2}log(r+\sqrt{r^2+\frac{k^2}{\lambda^2}})}}
\label{7.11} \\
\psi (r)&=& D_4\sqrt{D_3\mp \frac{|\lambda|r}{2}
\sqrt{r^2+\frac{k^2}{\lambda^2}} \pm
\frac{|\lambda|k^2}{2\lambda^2}log(r+\sqrt{r^2+\frac{k^2}{\lambda^2}})}
\label{7.12}
\end{eqnarray}

where $D_3$ and $D_4$ are constants of integration. This gives for the
space-time
metric

\begin{eqnarray}
g_{tt}&=&-r^2 \label{7.13} \\
g_{rr}&=&\frac{r^2}{r^2+\frac{k^2}{\lambda^2}}(D_3\mp \frac{|\lambda|r}{2}
\sqrt{r^2+\frac{k^2}{\lambda^2}} \pm
\frac{|\lambda|k^2}{2\lambda^2}log(r+\sqrt{r^2+\frac{k^2}{\lambda^2}}))^{-1}
\nonumber \\
g_{\theta\theta}&=&(D_4)^2(D_3\mp \frac{|\lambda|r}{2}
\sqrt{r^2+\frac{k^2}{\lambda^2}} \pm
\frac{|\lambda|k^2}{2\lambda^2}log(r+\sqrt{r^2+\frac{k^2}{\lambda^2}}))
 \nonumber \end{eqnarray}

and doing a Taylor series expansion of the metric for $\frac{k}{\lambda}\ll r$
gives
to lowest order in $\frac{k}{\lambda}$, the magnetic solution to the
conventional
theory. Note here also that the physical magnetic field is the rescaled $B$:
$B_{phys}=\frac{B}{\sqrt{|\lambda}}$.

\section{Conclusions and outlook.}
The most interesting question to ask in a new physical theory, is if the
theoretical predictions given by the theory agrees with the known experimental
results. In the case of theories defined in space-time dimensions different
from
(3+1), we have of course no direct experimental results. This makes me asking
another
related question: What if we lived in (2+1)-dimensions and knew that our
space-time
was approximately Minkowskian, and that Yang-Mills equations described our
physics to
very high accuracy. Could we then rule out this unified theory, based on
experimental
results for Yang-Mills (Maxwell) theories? I would say that we could not. And
that is
because, as shown in both section 3 and 4, the equations of motion governing
the
matter Yang-Mills fields in the unified theory are the normal Yang-Mills
equations,
and that is what we can measure in experiments. It was also shown that for very
weak
Yang-Mills fields $A^{YM}_{phys}\ll \sqrt{\lambda}$ the unified theory agrees
with the
conventional coupling. It is first when we can measure the back-reaction of the
metric-field, that we can rule out one of the two proposed theories. All this
will be
of interest if the same construction, for the unified theory, will work equally
well
in (3+1)-dimensions. There are however other facts about the unified theory,
that could
discriminate it from being a good physical theory. One thing is that it could
be that
one gets non-causal propagation of the fields in a solution. I have not
addressed that
question in this paper.

If one now tries the same type of construction for a unified theory in
(3+1)-dimensions, one gets problem not encountered here for (2+1)-dimensions.
My
starting point for finding this unified theory was the Ashtekar Hamiltonian,
and the
possibility of generalizing it to arbitrary gauge groups. It was in
(3+1)-dimensions
that the generalization of Ashtekar's variables first was found \cite{PPgg}. It
was
there shown that there exist an infinite number of apparently distinct
generalizations,
while here for (2+1)-dimensions the generalization seems to be unique. This is
closely
related to the existence of an infinite number of cosmological constants in
(3+1)-dimensions \cite{Beng.Inf}, while there is only one cosmological constant
in
(2+1)-dimensions
\cite{PP2+1}. Disregarding that problem, there is another problem in
(3+1)-dimensions;
the Ashtekar formulation is complex, and one needs reality conditions to select
the
real physical solutions, which may be hard to find in a possible unified
theory.

The latest years there has been an increasing interest in (2+1)-dimensional
quantum
gravity. It would be interesting to see whether it is possible to quantize the
unified
theory also, using for instance the loop-representation quantization
\cite{Ash2+1}.
 In fact, my motivation to start looking for a unified theory in terms of
Ashtekar's
variables, was the possibility to find a theory that is better suited for the
loop-representation quantization in (3+1)-dimensions \cite{RS}, than the
conventional
Einstein-Yang-Mills theory \cite{ART}. See however Gambini and P\"{u}llin
\cite{GP}
for a treatment of the Einstein-Maxwell theory in the loop-representation.

Finally I summarize the most interesting features of the unified theory
presented in
this paper. For gauge-group $G^{tot}=SO(1,2)\otimes G^{YM}$, the equations of
motion
for the Yang-Mills part are the normal Yang-Mills equations. And for very weak
Yang-Mills fields, the unified theory agrees with the conventional
Einstein-Yang-Mills
theory to lowest order in the Yang-Mills fields.\\

\Large \underline{Acknowledgement}\normalsize\\

 I thank Ingemar Bengtsson for discussions, ideas and criticism throughout this
work.
I also thank Professor Abhay Ashtekar for helping me with a problem in an early
stage
of this work.

\end{document}